\documentclass[10pt,aps,prb,twocolumn,showpacs,amssymb,floatfix]{revtex4-1}
\usepackage{amsmath,graphics,epsfig,mathrsfs}
\usepackage{bm}
\usepackage{epigraph}

\newcommand{\etal}{{\em et al.~}}

\begin{document}
\title{Spin Transfer Torque with Spin Diffusion in Magnetic Tunnel Junctions}
\author{A. Manchon$^{1}$}
\email{aurelien.manchon@kaust.edu.sa}
\author{R. Matsumoto$^{2,3}$}
\author{H. Jaffres$^2$}
\author{J. Grollier$^2$}
\affiliation{$^1$ King Abdullah University of Science and Technology (KAUST), Physical Science and Engineering Division, Thuwal 23955-6900, Saudi Arabia;}
\affiliation{$^2$ Unit{\'e} Mixte de Physique CNRS/Thales and Universit{\'e} Paris Sud 11, Route D{\'e}partementale 128, 91767 Palaiseau, France;}
\affiliation{$^3$ National Institute of Advanced Industrial Science and Technology (AIST), Spintronics Research Center, Tsukuba, Ibaraki 305-8568, Japan}
\date{\today}

\begin{abstract}
Spin transport in magnetic tunnel junctions in the presence of spin diffusion is considered theoretically. Combining ballistic tunneling across the barrier and diffusive transport in the electrodes, we solve the spin dynamics equation in the metallic layers. We show that spin diffusion mixes the transverse spin current components and dramatically modifies the bias dependence of the effective spin transfer torque. This leads to a significant linear bias dependence of the out-of-plane torque, as well as a non-conventional thickness dependence of both spin torque components.
\end{abstract}
\maketitle

Current-driven control of the magnetization direction in magnetic nanodevices has been made possible by the prediction of Spin Transfer Torque (STT), by Slonczewski and Berger \cite{Slonc96}. The transfer of spin angular momentum between the spin-polarized electrical current and the local magnetization has been observed in various devices such as metallic spin-valves, magnetic tunnel junctions (MTJs) and magnetic domain walls \cite{review1,chapter}. A number of technological applications have been proposed, based on these devices, such as STT-MRAM, Race track memory etc. \cite{appli}.\par

The most promising candidate for memory applications to date is the MTJ \cite{huai} due to its high performances and good compatibility with the existing C-MOS technology \cite{ieee}. Therefore, understanding the nature of the spin transfer torque in such devices is of seminal importance. The first theoretical studies showed that in the ballistic limit of symmetric MTJs comprising semi-infinite ferromagnets, the spin torque is on the form \cite{theo,slonc07,manchon,xiao,wil,heiliger,tang}
\begin{eqnarray}\label{eq:1}
&&{\bf T}=T_\|{\bf M}\times({\bf P}\times{\bf M})+T_\bot{\bf M}\times{\bf P},\\
&&T_\|=a_1V+a_2V^2,\;T_\bot=b_0+b_2V^2.\label{eq:2}
\end{eqnarray}
The first term $T_\|$ in Eq. (\ref{eq:1}), referred to as the in-plane torque, competes with the magnetic damping allowing for self-sustained magnetic precessions and switching, whereas the second term $T_\bot$, referred to as the out-of-plane torque, acts like an effective field applied along ${\bf P}$. Here, ${\bf M}$ and ${\bf P}$ are the magnetization direction of the free and pinned layers, respectively and $V$ is the bias voltage applied across the junction. Interestingly, as shown in Eq. (\ref{eq:2}), the bias dependence of the spin torque components is well defined, the in-plane torque possessing both linear ($a_1$) and (small) quadratic components ($a_2\ll a_1$), and the out-of-plane torque being only quadratic ($b_2$), besides the zero-bias exchange coupling term ($b_0$). This bias dependence has been confirmed experimentally in FeCoB/MgO/FeCoB by spin-diode measurements \cite{sankey}. The quadratic bias dependence of the out-of-plane torque has serious implications on current-driven magnetization dynamics. Whereas the sign of the in-plane torque depends on the polarity of current injection, the out-of-plane torque is always in the same direction. Consequently, at positive polarity both torques favor the antiparallel configuration, whereas at negative polarity, the in-plane torque favors the parallel configuration while the out-of-plane torque favors the antiparallel one. This competition leads to back-hopping of the magnetization state which is detrimental for applications such as MRAM\cite{sun,oh}.\par

However, recent experiments have reported important discrepancies between the actual bias dependence of the spin torque and the one proposed in Eqs. (\ref{eq:1})-(\ref{eq:2}) \cite{deac,petit,li,sun,oh}. In particular, Oh {\em et al.} \cite{oh} showed that in an asymmetrically designed MTJ, the bias dependence of the out-of-plane torque acquires a linear contribution. This contribution reduces the competition between in-plane and out-of-plane torques for negative polarity and results in a net reduction of the back-hopping process. Several mechanisms have been proposed to alter the bias dependence of the spin torque, such as finite layer thickness\cite{xiao,wil,tang}, barrier and electrode asymmetry \cite{xiao,wil,tang,prb}, magnons \cite{li,magnon,prb} and spin-flip scattering in the electrodes \cite{li,prb}. Nevertheless, these mechanisms fail to explain the large linear bias dependence usually observed in the out-of-plane torque.\par

In the present study, we propose a model of spin transfer in a diffusive spin transport approach taking place in the metallic electrodes of MTJs. We will show that
the diffusion processes (i) mix the two spin components transverse to the local magnetization which has the result to (ii) dramatically modify the bias dependence of the
effective spin torque as well as its thickness dependence. In particular, we show that the out-of-plane torque acquires a significant linear bias dependent term. This result is obtained by solving the spin diffusion equation for the spin accumulation transverse vector in the metallic layers adjacent to the insulator and imposing the interfacial spin current as a boundary condition. In particular, our model adopts the point of view of Zhang {\em et al}. whereby magnon-assisted tunneling process, at high bias, quenches any transport of hot electrons over a short distance inside the ferromagnetic electrode resulting in a pure diffusive transport within the electrode \cite{zhang97}.

\begin{figure}[ht]
\centering
\includegraphics[width=8cm]{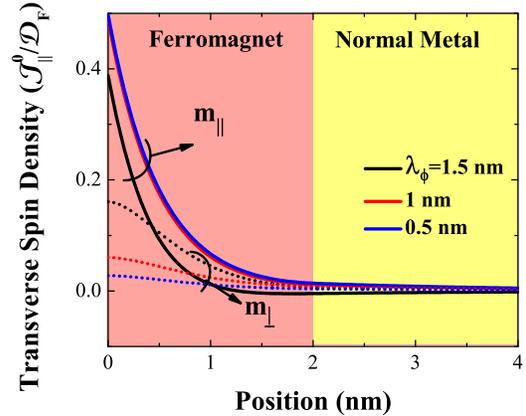}
\caption{\label{fig:Fig1}(Color online) Transverse spin accumulation as a function of the distance in the electrode in F(2 nm)/N. The parameters are $\lambda_{J}$=1 nm,$\lambda_{sf}^F$=15 nm, $\lambda_{sf}^N$=2 nm and ${\cal D}_F/{\cal D}_N=1$.}
\end{figure}

Let us consider a MTJ composed of N/F/I/F/N stack, where I is the insulating spacer, F are the ferromagnetic layers and N the non magnetic metallic leads connected to the F layer. The tunneling process through the insulating barrier imposes a ballistic injection of carriers at I/F interface which can be described by several calculation techniques like the free electron model, a tight-binding treatment or a density functional theory. These different approaches are well documented in a recent review paper Ref.~\onlinecite{review}. From the above arguments, it results that the two components of the transverse spin current $\mathcal{J}_0$ are imposed at the I/F interface. It constitutes a {\em boundary condition} to the coupled diffusive spin transport-relaxation equations for the transverse component of the spin accumulation vector $\bf{m}$. Along these guidelines, the spin dynamics of the transverse spin accumulation in the ferromagnetic electrode from the I/F interface is governed by the following time-dependent coupled equations
\begin{eqnarray}
\frac{\partial {\bf m}}{\partial t}&=&-{\bm \nabla}\cdot{\cal J}-\frac{1}{\tau_J}{\bf m}\times{\bf M}-\frac{1}{\tau_\phi}{\bf M}\times({\bf m}\times{\bf M})-\frac{{\bf m}}{\tau_{sf}},\label{eq:s0}\\
{\cal J}&=&-{\cal D}{\bm\nabla}\otimes{\bf m},\label{eq:s01}
\end{eqnarray}
where ${\bf m}$ is the spin accumulation, ${\bf M}$ is the direction of the localized \textit{3d} magnetization, $t$ is the time, and ${\cal J}$ is the spin current tensor. In Eq. (\ref{eq:s01}), the expression of the spin current is limited to Ohm's law \textit{via} the diffusion constant $\cal{D}$. These equations account for spatial variation of spin current, spin precession, spin dephasing and spin relaxation through the respective spin precession time $\tau_J$, spin decoherence time $\tau_{\phi}$ and spin relaxation time $\tau_{sf}$. Whereas the spin relaxation affects the three spin components, the spin precession and spin decoherence terms only affect the two transverse components of the spin accumulation vector. In the transient regime of spin injection dynamics, these equations can be solved analytically using Green's functions techniques in presence of a source term ${\cal J}_0 \delta(z)\delta(t)$ ($\delta$ is the Dirac distribution and $z$ is the direction of the tunneling current where $z=0$ correspond to the exact I/F interface position). In the steady-state regime of spin injection ($\times$ $\int_0 ^{\infty} dt$), it simply gives
\begin{eqnarray}
{\bm \nabla}\cdot{\cal J}&=&-\frac{1}{\tau_J}{\bf m}\times{\bf M}-\frac{1}{\tau_\phi}{\bf M}\times({\bf m}\times{\bf M})-\frac{1}{\tau_{sf}}{\bf m},\label{eq:s1}\\
{\cal J}&=&-{\cal D}{\bm\nabla}\otimes{\bf m},\label{eq:s2}
\end{eqnarray}
The solutions for the spin accumulation in the ferromagnet, when the magnetization is along ${\bf z}$ reads
\begin{eqnarray}
m_t&=&m_\|+i m_\bot=Ae^{x/L}+Be^{-x/L}\label{eq:mt}\\
\frac{1}{L^2}&=&-\frac{i}{\lambda_J^{2}}+\frac{1}{\lambda_\phi^2}+\frac{1}{\lambda_{sf}^2}\label{eq:L}
\end{eqnarray}
where $\lambda_i^2={\cal D}\tau_i$. We are now interested in the spin accumulation in the ferromagnetic layer, displayed in Fig. \ref{fig:Fig1}. On the left interface I/F, the boundary conditions are given by the interfacial spin current injected through the barrier ${\cal J}_0={\cal J}^0_\|+i{\cal J}^0_\bot$. The two components are the in-plane and out-of-plane tunneling spin currents defined by ${\cal J}_0={\cal J}^0_\|{\bf M}\times({\bf P}\times{\bf M})+{\cal J}^0_\bot{\bf M}\times{\bf P}$. On the right interface F/N, the connection is given by the continuity of the spin accumulation and spin current. In principle, interfacial resistance and spin-flip give additional contributions to the interfacial boundary conditions. In the present model, we disregard these effects since they can be added by simply inserting an effective interfacial layer. After some algebra, the two components of the transverse spin density read
\begin{eqnarray}
&&m_\|+im_\bot=\frac{{\cal J}_0L}{{\cal D}_F}\frac{\sinh\frac{d-x}{L}+\eta\cosh\frac{d-x}{L}}{\cosh\frac{d}{L}+\eta\sinh\frac{d}{L}},\;x<d\\
&&m_\|+im_\bot=\frac{{\cal J}_0L}{{\cal D}_F}\frac{\eta e^{-\frac{x-d}{\lambda_sf}}}{\cosh\frac{d}{L}+\eta\sinh\frac{d}{L}},\;x>d
\end{eqnarray}
where $\eta=\frac{{\cal D}_F}{{\cal D}_N}\frac{\lambda_{sf}^N}{L}$, ${\cal D}_i$ is the conductivity of the $i$-th layer and $d$ is the thickness of the ferromagnetic layer. Figure \ref{fig:Fig1} displays the spatial profile of the spin accumulation $m_\|$ and $m_\bot$ transverse to the local magnetization for different spin dephasing length $\lambda_\phi$=0.5 nm, 1 nm, and 1.5 nm. In this figure, we assumed ${\cal J}_\bot^0$=0 (no ballistic out-of-plane spin current). For the chosen parameters, the spin accumulation is not fully absorbed in the ferromagnet and decays away from the I/F interface. Furthermore, due to the spin precession, the spin accumulation displays both in-plane and out-of-plane components.\par
Note that the out-of-plane component $m_\bot$ increases when enhancing the spin dephasing length. Therefore, even in the absence of out-of-plane interfacial spin current (${\cal J}_\bot^0$=0), the interplay between spin dynamics and spin diffusion is expected to produce a torque with both in-plane and out-of-plane components. The spin torque is defined as the spatial change of spin current, compensated by the spin relaxation term
\begin{eqnarray}
&&{\bf T}=\frac{1}{\Omega}\int_{\Omega}d\Omega\left(-{\bm \nabla}\cdot{\cal J}-\frac{1}{\tau_{sf}}{\bf m}\right).
\end{eqnarray}
Here, $\Omega$ is the volume of the magnetic layer. Therefore, the total spin torque exerted on the ferromagnet is
\begin{eqnarray}
T_\|+iT_\bot=\frac{{\cal J}_{0}}{d}\frac{L^2}{L_0^2}\frac{\cosh\frac{d}{L}+\eta\sinh\frac{d}{L}-1}{\cosh\frac{d}{L}+\eta\sinh\frac{d}{L}},\label{eq:12}\end{eqnarray}
where $\frac{1}{L_0^2}=-\frac{i}{\lambda_J^{2}}+\frac{1}{\lambda_\phi^2}$.

In the limit of infinite ferromagnetic layer thickness, the torque in Eq. (\ref{eq:12}) reduces to
\begin{eqnarray}\label{eq:tx}
&&T_\|=\frac{1}{d\xi}\left(((1+\chi)\beta^2+\chi^2){\cal J}_{\|}^0+\chi^2\beta{\cal J}_{\bot}^0\right),\\
&&T_\bot=\frac{1}{d\xi}\left(((1+\chi)\beta^2+\chi^2){\cal J}_{\bot}^0-\chi^2\beta{\cal J}_{\|}^0\right),
\label{eq:ty}
\end{eqnarray}
where $\chi=\tau_\phi/\tau_{sf}$, $\beta=\tau_J/\tau_{sf}$ and $\xi=(1+\chi)^2\beta^2+\chi^2$. In the case of very short spin dephasing $\chi\ll1$, as in Fe/MgO/Fe\cite{heiliger}, the spin torque reduces to the ballistic limit, $T_\|={\cal J}_{\|}^0/d$ and $T_\bot={\cal J}_{\bot}^0/d$. In the limit of infinite spin dephasing $\chi\gg1$, such as the one considered in Ref. \onlinecite{li,prb}, the spin torque arises from a mixture of ${\cal J}_{\|}^0$ and ${\cal J}_{\bot}^0$, $T_\|=({\cal J}_{\|}^0+\beta{\cal J}_{\bot}^0)/d(1+\beta^2)$ and $T_\bot=({\cal J}_{\bot}^0-\beta{\cal J}_{\|}^0)/d(1+\beta^2)$. Therefore, in the case of semi-infinite ferromagnets, a linear bias dependence of the out-of-plane torque arises only if both the spin dephasing and the spin precession are comparable to the spin diffusion length.\par

\begin{figure}
\centering
\includegraphics[width=8cm]{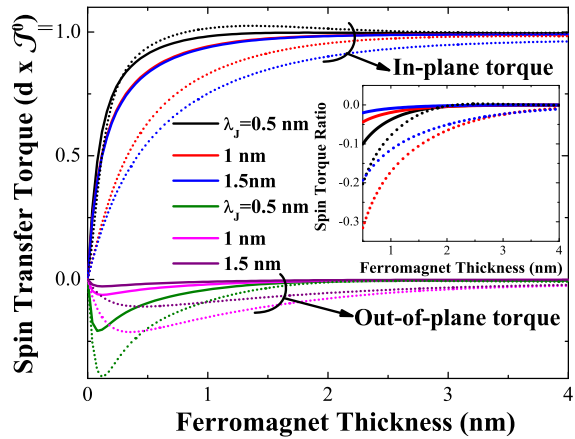}
\caption{\label{fig:Fig2}(Color online) Thickness dependence of the normalized spin torques $dT_\|$ and $dT_\bot$ for different spin precession lengths, $\lambda_J=0.5$ nm, 1 nm and 1.5 nm. The parameters are the same as in Fig. \ref{fig:Fig1}, except for the spin dephasing length: $\lambda_\phi=0.5$ nm (solid lines) and $\lambda_\phi$=1 nm (dotted lines). Inset: Thickness dependence of the ratio between out-of-plane torque and in-plane torque for the same parameters.}
\end{figure}

The case of a finite ferromagnetic layer presents significant differences due to the adjacent normal metal. When the spin precession and dephasing occur on a length much smaller than the thickness of the ferromagnet, the transverse spin current is mostly absorbed at the interface between the insulator and the ferromagnet and the spin torque bias dependence is described by Eqs. (\ref{eq:1})-(\ref{eq:2}). Conversely, when the spin dephasing and precession extend on a length comparable to the layer thickness, the transverse spin current is no more confined to the interface and extends on the layer thickness to the second interface with the normal metal, as shown in Fig. \ref{fig:Fig1}. This new dynamics mixes the two transverse components of the spin current and is therefore responsible for the deviation from the ballistic bias dependence shown in Eq. (\ref{eq:2}). Figure \ref{fig:Fig2} displays the two components of the spin torque, $T_\|$ and $T_\bot$, obtained from Eq. (\ref{eq:12}) in the case ${\cal J}_{\bot}^0$=0 as a function of the thickness of the ferromagnetic layer. In this figure, the two contributions have been multiplied by the distance $d$ to remove the $1/d$ dependence coming from the interfacial nature of the spin torque. Interestingly, it appears that both in-plane and out-of-plane torques possess a thickness-dependence that is not simply $\propto1/d$, as it is expected in the case of purely interfacial spin torque.\par

A deviation from the 1/$d$-thickness dependence of the spin torque has been identified previously in calculations assuming a ballistic regime \cite{heiliger,wil}. For example, Ref. \onlinecite{heiliger} observes a deviation that is attributed to resonant quantum states in the finite ferromagnetic layer. Note that the model proposed in the present work assumes a diffusive transport in the metallic layers, i.e. defects and disorder are strong enough to quench quantum coherence. Therefore, the deviation we observe here is not related to quantum states, but rather to the incomplete absorption of the spin current: when the thickness of the free layer is on the order of or smaller than the spin dephasing length $\lambda_\phi$, the transverse spin current responsible for the spin torque is not fully absorbed in the free layer and diffuses towards the capping layer. This induces a deviation from the usual 1/$d$-thickness dependence. Increasing the thickness of the free layer improves the absorption of the spin current and for thicknesses much larger than the spin dephasing length, the thickness dependence of the torque recovers the 1/$d$ limit (see Fig. \ref{fig:Fig2}). Note that in the case of half-metallic behavior, as in Fe/MgO/Fe tunnel junctions, the minority band with $\Delta_1$ symmetry does not propagate in the ferromagnet which results in a quenching of the spin dephasing length ($d\gg \lambda_\phi$) and reduces the spin torque to a 1/$d$ behavior.

Since the transverse spin accumulation is not fully absorbed in the ferromagnet, the remaining unabsorbed transverse spin accumulation diffuses into the normal metal and modifies the actual spin torque exerted on the ferromagnet. More specifically, the mixing that gives rise to both components increases for smaller thicknesses. Therefore, the ratio between the out-of-plane and the in-plane components, $T_\bot/T_\|$, increases when decreasing the thickness of the ferromagnet, as displayed in the inset in Fig. \ref{fig:Fig2}. The nature of the spin torque asymptotically tends towards the semi-infinite ferromagnet limit for thicknesses much larger than the spin dephasing length.\par

Let us now illustrate the influence of the normal metal on the spin diffusion in the ferromagnetic layer. The role of the normal metal can be qualitatively understood if one models the F/N bilayer by an equivalent semi-infinite F layer with an effective spin diffusion length  $\lambda^*_{sf}$. In the semi-infinite limit, the spin torque reduces to
\begin{eqnarray}
T_\|+iT_\bot&\xrightarrow{d\rightarrow\infty}&\frac{{\cal J}_{0}}{d}\frac{L^{*2}}{L_0^2},\label{eq:23}
\end{eqnarray}
where $\frac{1}{L^{*2}}=-\frac{i}{\lambda_J^{2}}+\frac{1}{\lambda_\phi^2}+\frac{1}{\lambda_{sf}^{*2}}$, $\lambda^*_{sf}$ being the effective spin diffusion length arising from the presence of the adjacent normal metal. By equating Eq. (\ref{eq:23}) with Eq. (\ref{eq:12}), one can define the effective spin diffusion length $\lambda^*_{sf}$ as a function of the parameters of the normal layer. To obtain a tractable analytical result, we consider an ultrathin ferromagnet, so that
\begin{eqnarray}
T_\|+iT_\bot&\xrightarrow{d\rightarrow 0}&\frac{{\cal J}_{0}}{d}\frac{L^{2}}{L_0^2}\frac{1}{1+\frac{L}{\eta d}}.\label{eq:24}
\end{eqnarray}
This provides the analytical expression
\begin{eqnarray}
\frac{1}{L^{*2}}=\frac{1}{L^2}+\frac{{\cal D}_N}{{\cal D}_F}\frac{1}{\lambda^N_{sf}d},\end{eqnarray}
or equivalently
\begin{eqnarray}
&&\frac{1}{\tau_{sf}^{F*}}=\frac{1}{\tau_{sf}^F}+\frac{1}{\tau_{sf}^N}\frac{\lambda^N_{sf}}{d}=\frac{1}{\tau_{sf}^F}+\frac{p_N}{1-p_N}\frac{1}{\tau_{sf}^N},\label{eq:tsf}\end{eqnarray}
where $p_N$ is the probability to find the particle in N. The bulk spin relaxation time in the finite ferromagnet $\tau_{sf}^F$ is renormalized by the presence of the normal metal through the probability $p_N$. Increasing the spin diffusion length of the normal metal $\lambda^N_{sf}$ or decreasing the thickness of the ferromagnet $d$ reduces the spin diffusion length in the ferromagnet. The second term of Eq. (\ref{eq:tsf}) can be viewed as the effective spin lifetime in the finite ferromagnetic layer.\par

\begin{figure}
\centering
\includegraphics[width=9cm]{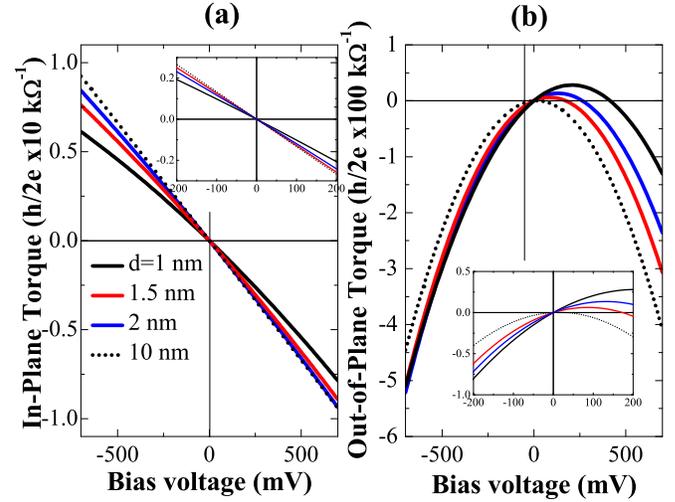}
\caption{\label{fig:Fig3}(Color online) Bias dependence of the in-plane (a) and out-of-plane torques (b) for different thicknesses of the ferromagnetic layer. The parameters are $\lambda_J=\lambda_\phi=$1 nm, $\lambda_{sf}^F$=10 nm, $\lambda_{sf}^N$=1 nm and ${\cal D}_F/{\cal D}_N$=1. The bias dependence of the interfacial spin current ${\cal J}_0$ is taken from the spin torque bias dependence measured by Sankey \etal \cite{sankey}. The inset display the zoom in the range $\pm200$ mV.}
\end{figure}

This thickness dependence has a very important implication on the bias dependence of the spin torque. If one assumes a bias dependence of the spin current on the form ${\cal J}_{\|}^0=a_1V$ and ${\cal J}_{\bot}^0=b_2V^2$, as expected and observed for systems such as Fe/MgO/Fe \cite{sankey,heiliger}, then both in-plane and out-of plane spin torque components will be a mixture of linear and quadratic bias dependences. For example, Sankey \etal and Kubota \etal experimentally found that $T_\bot/T_\||_{200mV}\approx$ 10\% to 15\% \cite{sankey}. Assuming that the bias dependence of the interfacial spin current ${\cal J}_0$ is given by the spin torque bias dependence measured by Sankey \etal, the in-plane and out-of-plane torques are reproduced in Fig. \ref{fig:Fig3} for different thicknesses of the ferromagnetic layer. As expected from the discussion above, the linear character increases when decreasing the ferromagnetic thickness and at large thicknesses it reduces to the bulk value given by Eq. (\ref{eq:ty}). Since the ballistic out-of-plane spin current ${\cal J}_{\bot}^0$ is small compared to the in-plane spin current ${\cal J}_{\|}^0$, only slight modification of the in-plane torque is expected.\par

In conclusion, the influence of spin diffusion in the metallic layers of MTJs on the spin transfer torque has been addressed theoretically. Assuming an interfacial bias-driven spin current at the interface between the insulator and the ferromagnet, the spin diffusion equation is solved and describes a complex spin dynamics in the metallic layers. It is found that this dynamics mixes the components of the spin current tranverse to the local magnetization which results in a superposition between linear and quadratic bias dependence for both the in-plane and out-of-plane torques. The thickness dependence of the spin transfer torque is also altered for small thicknesses.\par

The authors acknowledge fruitful discussions with A. Fert.


\begin{thebibliography} {999}
\bibitem{Slonc96} J. C. Slonczewski, J. Magn. Magn. Mater. {\bf159}, L1 (1996); L. Berger, Phys. Rev. B {\bf54} 9353, (1996).
\bibitem{review1} D. C. Ralph and M. D. Stiles, J. Magn. Magn. Mater. {\bf 320}, 1190�1216 (2008); J. Z. Sun and D. C. Ralph, J. Magn. Magn. Mater. {\bf 320}, 1227 (2008).
\bibitem{chapter} A. Manchon, and S. Zhang, 'Spin Torque in Magnetic Systems: Theory', Handbook of Spin Transport and
Magnetism, Eds. E.-Y. Tsymbal and I. Zutic, Chap. 8, CRC Press, August 2011.
\bibitem{appli} C. Chappert, A. Fert and F. Nguyen Van Dau, Nature Materials {\bf6}, 813 (2007); S. S. P. Parkin \etal, Science {\bf320}, 190 (2008).
\bibitem{huai} J.Z. Sun, J. Magn. Magn. Mater. {\bf202}, 157 (1999); Y. Huai \etal, Appl. Phys. Lett. {\bf84}, 3118
(2004); G. D. Fuchs \etal, Appl. Phys. Lett. {\bf85}, 1205 (2004); D. Chiba \etal, Phys. Rev. Lett. {\bf93}, 216602 (2004).
\bibitem{ieee} S. Ikeda, J. Hayakawa, Y. M. Lee, F. Matsukura, Y. Ohno,
T. Hanyu, and H. Ohno, IEEE Trans. Elec. Dev. {\bf54}, 991 (2007).
\bibitem{theo} I. Theodonis \etal, Phys.Rev. Lett. {\bf97}, 237205 (2006).
\bibitem{slonc07} J. C. Slonczewski, Phys. Rev. B {\bf 71}, 024411 (2005); J.C. Slonczewski and J.Z. Sun, J. Magn. Magn. Mater.
{\bf310}, 169-175 (2007); See also, J. C. Slonczewski, Phys. Rev. B {\bf 39}, 6995 (1989).
\bibitem{manchon} A. Manchon \etal, J. Phys.: Condens. Matter {\bf20}, 145208 (2008); {\em ibid} {\bf19}, 165212 (2007).
\bibitem{xiao} J. Xiao, G. E. W. Bauer, and A. Brataas, Phys. Rev. B {\bf77}, 224419 (2008).
\bibitem{wil} M. Wilczynski, J. Barnas, and R. Swirkowicz, Phys. Rev. B {\bf77}, 054434 (2008).
\bibitem{heiliger} C. Heiliger and M. D. Stiles, Phys. Rev. Lett. {\bf100}, 186805 (2008).
\bibitem{tang} Y.-H. Tang \etal, Phys. Rev. Lett. {\bf103}, 057206 (2009); Phys. Rev. B {\bf81}, 054437 (2010).
\bibitem{sankey} J. C. Sankey \etal, Nature Physics {\bf4}, 67
(2008); H. Kubota \etal, Nature Physics {\bf4}, 37 (2008).
\bibitem{sun}J. Z. Sun \etal, J. Appl. Phys. {\bf105}, 07D109 (2009); T. Min \etal, J. Appl. Phys. {\bf105}, 07D126 (2009).
\bibitem{oh} S.-C. Oh \etal,, Nature Physics {\bf 5}, 898 (2009).
\bibitem{deac} A. M. Deac \etal, Nature Physics {\bf4}, 803 (2008).
\bibitem{petit}S. Petit \etal, Phys. Rev. Lett. {\bf98}, 077203 (2007).
\bibitem{prb} A. Manchon, S. Zhang and K.-J. Lee, Phys. Rev. B {\bf82}, 174420 (2010).
\bibitem{li} Z. Li \etal, Phys. Rev. Lett. {\bf100}, 246602 (2008).
\bibitem{magnon} P. M. Levy and A. Fert, Phys. Rev. Lett. {\bf97}, 097205 (2006); Phys.Rev. B {\bf74}, 224446 (2006); A. Manchon and S. Zhang, Phys. Rev. B {\bf79}, 174401 (2009).
\bibitem{zhang97} S. Zhang, P. M. Levy, A. C. Marley, and S. S. P. Parkin, Phys. Rev. Lett. {\bf79}, 3744 (1997).
\bibitem{review} K. D. Belashchenko, and E.-Y. Tsymbal, 'Tunneling Magnetoresistance: Theory', Handbook of Spin Transport and
Magnetism, Eds. E.-Y. Tsymbal and I. Zutic, Chap. 12, CRC Press, August 2011.
\end{thebibliography}
\end{document}